# Diagram for vortex formation in quasi-two-dimensional magnetic dots

J. C. S. Rocha,[1,a)] P. Z. Coura,[2,b)] S. A. Leonel,[2,b)] R. A. Dias,[2,b)] and B. V. Costa[1,a)]
[1]*Departamento de Física, Laboratório de Simulação, ICEX, UFMG, 30123-970 Belo Horizonte, MG, Brazil*
[2]*Departamento de Física, ICE, UFJF, 36036-330 Juiz de Fora, MG, Brazil*

The existence of nonlinear objects of the vortex type in two-dimensional magnetic systems presents itself as one of the most promising candidates for the construction of nanodevices, useful for storing data, and for the construction of reading and writing magnetic heads. The vortex appears as the ground state of a magnetic nanodisk whose magnetic moments interact via the dipole-dipole potential $\{D\Sigma[\vec{S}_i \cdot \vec{S}_j - 3(\vec{S}_i \cdot \hat{r}_{ij}) \times (\vec{S}_j \cdot \hat{r}_{ij})]/r_{ij}^3\}$ and the exchange interaction $(-J\Sigma \vec{S}_i \cdot \vec{S}_j)$. In this work it is investigated the conditions for the formation of vortices in nanodisks in triangular, square, and hexagonal lattices as a function of the size of the lattice and of the strength of the dipole interaction $D$. Our results show that there is a "transition" line separating the vortex state from a capacitorlike state. This line has a finite size scaling form depending on the size, $L$, of the system as $D_c = D_0 + 1/A(1+BL^2)$. This behavior is obeyed by the three types of lattices. Inside the vortex phase it is possible to identify two types of vortices separated by a constant, $D = D_c$, line: An in-plane and an out-of-plane vortex. We observed that the out-of-plane phase does not appear for the triangular lattice. In a two layer system the extra layer of dipoles works as an effective out-of-plane anisotropy inducing a large $S^z$ component at the center of the vortex. Also, we analyzed the mechanism for switching the out-of-plane vortex component. Contrary to some reported results, we found evidences that the mechanism is not a creation-annihilation vortex anti-vortex process. © *2010 American Institute of Physics*. [doi:10.1063/1.3318605]

## I. INTRODUCTION

The miniaturization of electronic devices has a natural limit imposed by the thermal fluctuations which determines how long the magnetization of a ferromagnetic structure survives, or in other words: The long-range ferromagnetic order vanishes when the energy due to the anisotropy becomes comparable to the thermal fluctuation energy in the system. That is the well known superparamagnetic limit[1] that impose physical limits in the miniaturization of magnetic devices. Recent developments in nanomagnetic materials has shown that the development of a vortex in quasi-two-dimensional (2D) nanomagnets can help to overcome the superparamagnetic limit. Their expected applications include magnetic random access memory, high density magnetic recording media, magnetic sensors, and magnetic reading and writing heads.[2–4]

By a vortex we mean a special configuration of magnetic moments similar to the stream lines of a circulating flow in a fluid. The magnetic moments precess by $\pm 2\pi$ on a closed path around the vortex. In the Fig. 1 we show schematically the types of vortices and antivortices that can appear in magnetic systems. The importance of vortices in magnetic systems is known since the early seventies in connection with the Berezinskii–Kosterlitz–Thouless (BKT) phase "transition." In a seminal work Berezinskii[5] and later Kosterlitz and Thouless[6] showed that the easy plane Heisenberg model (EPHM) in two dimensions undergoes an infinite order phase transition. The EPHM is described by the Hamiltonian $H = \Sigma_{<i,j>} -J\vec{S}_i \cdot \vec{S}_j + A\vec{S}_i^z \cdot \vec{S}_j^z$, where $J$ is an exchange term and $A$ an easy plane anisotropy. The EPHM has a BKT transition at a temperature $T_{\rm BKT}$ coming from a high-temperature phase where the same time space-space correlation function exhibits an exponential decay to a low-temperature phase with

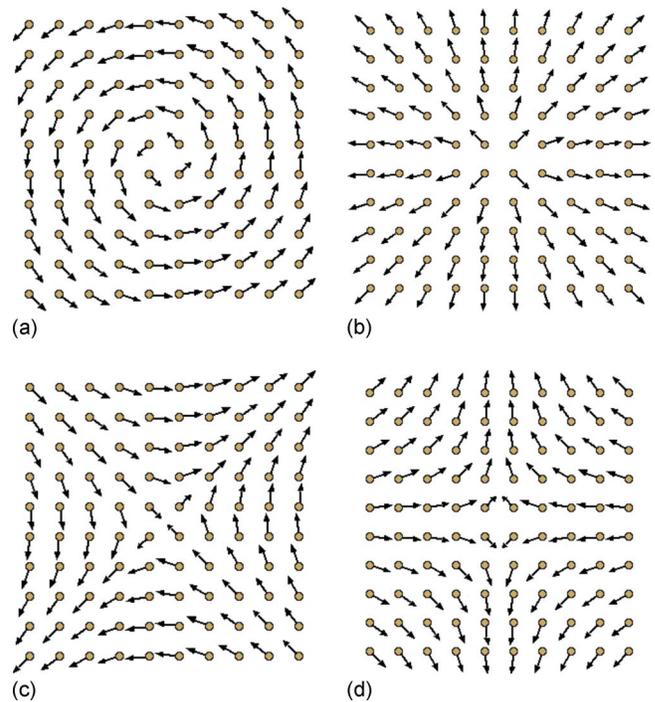

FIG. 1. (Color online) Show schematically in a square lattice: (a) type I vortex, (b) type I antivortex, (c) type II vortex, and (d) type II antivortex.

[a)]Electronic addresses: jcsrocha@fisica.ufmg.br and bvc@fisica.ufmg.br.
[b)]Electronic addresses: pablo@fisica.ufjf.br, sidiney@fisica.ufjf.br, and radias@fisica.ufjf.br.

quasi-long-range order where the correlation function has a power-law decay. This phase transition is believed to be driven by a vortex-antivortex unbinding mechanism. The energy associated with the pair vortex–antivortex is proportional to ln $r_{v-av}$, where $r_{v-av}$ is the distance between the vortex and antivortex centers. The logarithmic behavior of the energy prohibits the existence of an odd number of vortices or antivortices in the EPHM with free or periodic boundary conditions. That is because an isolated vortex (or antivortex) has energy ln $R$, where $R$ is the vortex size. In a magnetic nanodot a magnetic dipole-dipole energy term has to be considered beside the exchange and the anisotropy terms.[7] The dipole energy competes with the exchange term so that for large enough dipole interaction the continuity of the magnetic field in the boundary of the system imposes the magnetic moments to be tangent to the border of the nanodot.[8–10] This kind of boundary condition favors the appearing of an isolated vortex at the center of the system (See Fig. 1). Although, the vortices and antivortices shown in Fig. 1 have the same bulk energy, competition with the border energy clearly favors the appearing of the vortex of type *I* in the system. Due to the singularity at the vortex core the system can lower its energy by developing an out-of-plane magnetization, perpendicular to the plane of the disk (the *z* direction).[11–14] Experimental observations in circular dots of Permalloy suggest that states "up" and "down" ($\pm z$) are degenerated not depending on the vortex orientation (clockwise or counterclockwise).[15] The existence of this degeneracy suggests that it is possible to store one bit of information in this degree of freedom.[16–18] Experimental results also show that the effective anisotropy is very small and can be neglected[15] in theoretical modeling. Because the vortex formation energy is proportional to ln $R$ the superparamagnetic limit is pushed down opening the possibility of building smaller magnetic devices for storing data than those allowed by the nowadays technology.[7]

In this work we report a study of quasi-2D magnetic dots by using Monte Carlo (MC) and spin dynamics simulations.[19] The system is modeled by distributing magnetic particles over a lattice. The particles interact through exchange and dipolar potentials. We study the vortex formation in three different types of lattices: hexagonal, square, and triangular. Also, we discuss the stability of the vortex as a function of the strength of the dipole interaction and the system size. The paper is organized as follows. In Sec. II we present the model we will deal with and some considerations about the simulations. In Sec. III we present our results and discussions. In Sec. IV our conclusions are presented.

## II. MODEL

Theoretically we can write a model Hamiltonian for a magnetic nanodot in a pseudospin language as[7]

$$H = -J \sum_{\langle i,j \rangle} \vec{S}_i \cdot \vec{S}_j + D \sum_{i \neq j} \left[ \frac{\vec{S}_i \cdot \vec{S}_j}{r_{ij}^3} - \frac{3(\vec{S}_i \cdot \vec{r}_{ij}) \times (\vec{S}_j \cdot \vec{r}_{ij})}{r_{ij}^5} \right]. \quad (1)$$

Here *J* is an exchange coupling constant, $\vec{S}_i$ and $\vec{S}_j$ are spin variables defined on sites *i* and *j*, $r_{i,j}$ is the distance between spins at *i* and *j*, and *D* is the dipole strength. The sum in the first term is over first neighbors and the sum in the second term considers a cut-off in the dipolar interaction up to a neighbor $r_{ij} \leq r_{\text{cut}}$. As will be discussed in the following our results showed that the cut-off in the dipolar interaction has to be taken very carefully. We can understand the physical model described by Hamiltonian Eq. (1) as follows: the first term (the exchange interaction) tends to align the spins of neighboring sites. The second term (the dipole interaction) is divided in two parts: the first one tends to align the spins antiferromagnetically. The second part tends to align the spins along the direction of the unity vector that connects the sites *i* and *j*. At the border of the system the magnetic moments are aligned tangent to the border satisfying both, the condition that minimizes the exchange interaction and the second part of the dipole energy term. In a vortex configuration the first part of the dipole interaction is only minimized for spins at sites in opposing positions in relation to the center of the vortex.

The dipole interaction is in general very difficult to treat in any analytical or computational calculation due to its long range character. Several works[8–10] that deal with magnetic nanodots use a variation in the Hamiltonian Eq. (1) by considering an anisotropic interaction $\Sigma(\vec{S}_i \cdot \vec{n}_i)^2$ to replace the dipole term. Here, $\vec{n}_i$ represents a unit vector perpendicular to the surface and to the borderline of the system. This term contributes positively to the total energy, therefore, it forces the spins to be perpendicular to $\vec{n}_i$, competing with the exchange term. The effect produced by this anisotropic interaction is similar to that one of the dipole term. It favors the magnetic moment into a configuration tangent to the border of the system and parallel to the disk plane. The energy due to this term is minimized when the magnetic moments arrange themselves in a curling vortex structure. The low temperature properties of the Hamiltonian with the anisotropic term is similar to the one obtained by using the long range dipole interaction. However, the high temperature and the dynamical behaviors are quite different. As we want to explore the model beyond its low temperature properties, we treat the system by using Eq. (1). So far, much of the theoretical and computational work done to understand the behavior of vortices in magnetic nanosystems consider a 2D model. Although, the results obtained by using this simplification are in quite good agreement with experimental findings, part of the present work is dedicated to discuss the influence of an additional layer in the 2D model. The nanodot is defined as follows. An number of magnetic particles is distributed over the lattice points of a 2D array. A circle of size *L* centered in a previously chosen cell is drawn over the array. Here *L* is measured in units of the lattice parameter *a*. The sites outside the circle are erased. The sites left inside form the nanodot, that is the object of our interest. If we are interested in a two layer nanodot the building process is similar. In this case the nanodot will be a small cylinder of diameter *L* and height *a*. As a matter of simplification from now on distances are measured in units of *a*, defined as the distance between first neighbors sites in the lattice.

A numerical approach to study this model is always very time consuming since we must consider the interactions of each spin on a site $i$ with all others in the system. It makes the computer time prohibitive for large values of $L$. In order to reduce the computational time much of the work done so far considers a cut-off in the dipolar interaction up to a neighbor $r_{cut}$. However, the introduction of a cut-off creates distortions in the ground state of the system as will be discussed below. In order to understand the effect of the cut-off we work the model defined by Hamiltonian Eq. (1) with and without a cut-off in the three different lattices: triangular, hexagonal, and square. For each of them we varied the dipole strength, $D/J$, in the range [0,0.50], for several sizes, $L$, with $10 < L < 90$. Without loss of generality we studied the triangular and the square lattices using only one layer, ($z=1$). Some exploratory results showed that the results for the two layer ($z=2$) system are similar to those of the hexagonal lattice.

To obtain the magnetic nanodot ground state of the model we used a numerical Metropolis MC method[20] combined with simulated annealing.[21] The simulated annealing approach is a generalization of the MC method to search for the ground state of a given system. We start with the system at a high temperature configuration. Then, a process of cooling is done slowly until a very low temperature is reached. As temperature decreases the magnetic moments in the system organizes themselves in a uniform structure with minimal energy. If the system is well behaved enough it is expected that the low temperature configuration approaches the ground state as close as we want. In our calculations we take the initial configuration at random, which corresponds to infinite temperature. The lower temperature is taken as $T=10^{-2}J/k_B$. From now on we consider $J=1$ and the Boltzman constant $k_B=1$ so that the energy and temperature are measured in units of $J$ and $J/K_B$, respectively. As a matter of comparison we choose in some cases the initial configuration as a vortex like structure. We found that the results for the ground state were quantitatively the same when compared with those obtained by using the disordered initial state. In our plots the error bars are smaller than the symbols when not shown. The results discussed in this work were obtained for the hexagonal lattice when not explicitly written.

## III. RESULTS AND DISCUSSION

In order to understand the influence of the cut-off and of the disk size in the ground state we simulated disks of several diameters ($10 \leq L \leq 90$) for different cutoffs. We observed that the most stable configuration can be a vortex or a capacitorlike structure (see Fig. 2), depending on the value of the dipole interaction, $D$. Our results are shown in Fig. 3 in a plot of $D_c$ as a function of the disk diameter for several values of $r_{cut}$. Here, $D_c$ represents the value of the dipole interaction at the crossing over value, when the ground state changes from the vortex to the capacitor configuration, labeled as *III* and *I*, respectively, in the figure. It is also shown a third state, labeled *II*, where the most stable vortex has an out-of-plane component. As a matter of clarity the region where the out-of-plane vortex appears is shown only for the simulation with no cut-off ($r_{cut}=L$). As can be seen in the

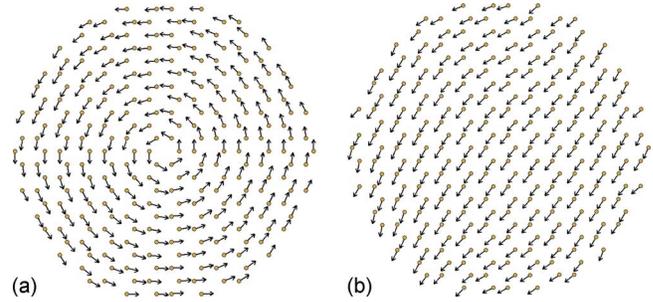

FIG. 2. (Color online) Schematic view of the vortex and capacitorlike configurations are shown in the left and right hand sides, respectively.

Fig. 3 the choice of the cut-off is important in the determination of the border between the vortex and capacitor regions. Because of that, we decided to use no cut-offs in our calculations even that implying in a longer CPU time for performing our calculations.

In Fig. 4 we plot the border lines defining the regions where the planar vortex, the out-of-plane vortex and capacitor configurations are more stable for all three types of lattices and $r_{cut}=L$. The figure can be understood as follows. Based in the model described by the Hamiltonian Eq. (1) it is possible to find a set of values of $D$ that minimizes the energy of the vortex configuration. To minimize the exchange interaction, at the center of the vortex, the magnetic moments tend to align in a direction perpendicular to the plane of the disk, which in turn maximizes the dipolar interaction. The vortex configurations of minimum energy can be in-plane or out-of-plane at the center of the vortex. This behavior depends on the size of the system $L$. The value of $D$ for vortex configurations with out-of-plane components decreases with increasing the value of $L$ because the contribution of the dipolar interaction (long-range) becomes greater than the exchange interaction.

In a two layer system, the additional layer acts as an anisotropy so that the net effect is similar to the introduction of an exchange anisotropy in the system.[15] It is well known that in an infinite system described by the anisotropic Heisenberg model in two dimensions, an out-of-plane exchange anisotropy can cause the effect of lowering the energy necessary to the vortex develop the $z$ component.[11,12] In a system with exchange anisotropy $A\sum s_i^z s_j^z$, there is a characteristic value of the exchange anisotropy, $A_c \approx 0.7$, at which the most stable vortex develops an out-of-plane component.[13,14] Because of that, we can expect that the out-of-plane phase appears at lower values of $L$ and higher $D/J$ when compared with the one layer case. In Fig. 5 we show the diagram for a $z=2$ system. The results confirm the expected picture. The diagram is similar to the one obtained for $z=1$ but with the out-of-plane phase starting at lower values of $L$ and higher $D/J$.

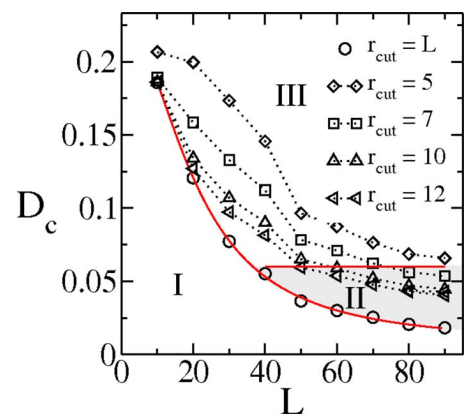

FIG. 3. (Color online) Diagram for the vortex formation in the case of the one layer hexagonal lattices. It is shown the critical values of the dipole interaction strength $D_c$, in units of $J$, as a function of the lattice size, $L$, for several values of the cut-off in the dipole interaction, $r_{cut}$. The circles correspond to the infinite range interaction. The lines separate two regions with different ground states. Regions I and III have a capacitor and a vortex (without out-of-plane component) in the ground state, respectively. As $r_{cut}$ increases the lines separating the phases approach the asymptotic line $r_\infty$. The shaded area represents a region (II) where the most stable configuration has an out-of-plane component at the center of the vortex.

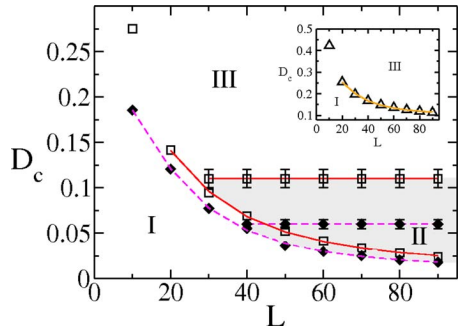

FIG. 4. (Color online) Diagram for the vortex formation in the three different types of the one layer lattices studied. Squares and diamonds correspond to the square and the hexagonal lattices, respectively. The inset shows the results for the triangular lattice. Region I and III have a capacitor and a vortex in the ground state, respectively. The shaded area (Region II) represents a region where the most stable configuration has an out-of-plane component at the center of the vortex. The lines separating the capacitor and the vortex states regions were adjusted using the equation $D_c = D_0 + 1/A(1+BL^2)$ with appropriate values of the constants $D_0$, $A$, and $B$, for each type of lattice.

We observed that in both cases, $z=1$ and 2, the capacitor and the vortex states regions are separated by a transition line that asymptotically tends to a constant, $(D_0)$. In any case this line can be very well adjusted by a function

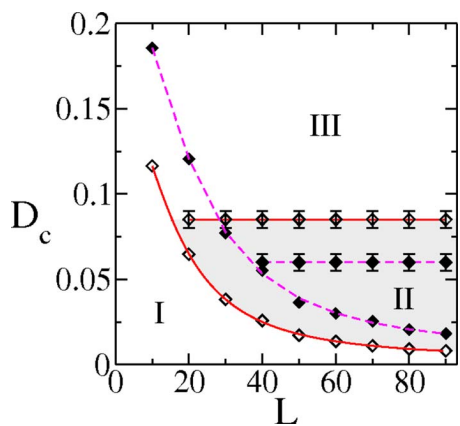

FIG. 5. (Color online) Diagram for the vortex formation in a two layer hexagonal lattice (open symbols) compared with the one layer hexagonal lattice (filled symbols) with $r_{cut}=L$. The regions I, II, and III are defined as in Figs. 3 and 4.

TABLE I. Parameters used in the Eq. (2).

|  | $D_0$ | $A$ | $B$ |
| --- | --- | --- | --- |
| Triangular | 0.098 | 3.798 | 0.002 |
| Squared | 0.011 | 4.417 | 0.002 |
| Hexagonal ($z=1$) | 0.007 | 4.522 | 0.002 |
| Hexagonal ($z=2$) | 0.003 | 6.375 | 0.004 |

$$D_c = D_0 + \frac{1}{A(1+BL^2)}. \qquad (2)$$

The values of the parameters are given in the Table I. Although, this is an *ad hoc* expression, a finite size scaling behavior of the boundary between the capacitor and vortex phases is clearly indicated. The constant $D=D_c$ line separates the in-plane and the out-of-plane vortex phases. We can describe this result as follows. Due to the competition between the dipolar interaction responsible for the formation of the vortex and the exchange ferromagnetic interaction, the system may develop a component in the $z$ direction, perpendicular to the plane of the vortex. The $z$ component is restrict to a region around the center of the vortex. For a small disk of diameter $L$, the influence of the edge is large dominating the behavior of the system. The energy due to a misalignment between the spins in the edge and in the center of the system is large enough to compete with the exchange energy. In such a case the vortex is expected to be planar. However, for even moderate disk sizes, the border plays no role. The out-of-plane core extends for only a few lattice constants allowing the spins to develop a $z$ component in the central region.

An important question is the thermodynamic behavior of the nanodot. As temperature increases, vortices, and antivortices are created in the system. They appear always as pairs. The energy associated with the pair excitation is given approximately by $\ln r_{v-av}$, where $r_{v-av}$ is the distance between the vortex and antivortex centers. In Fig. 6 we show two spin configurations in a hexagonal lattice of size $L=30$ for $D/J=0.1$ and $T/J=0.7$. Figure 7 shows the vortex density as a function of temperature. We observe that there is a threshold below which there are no pairs present in the system.

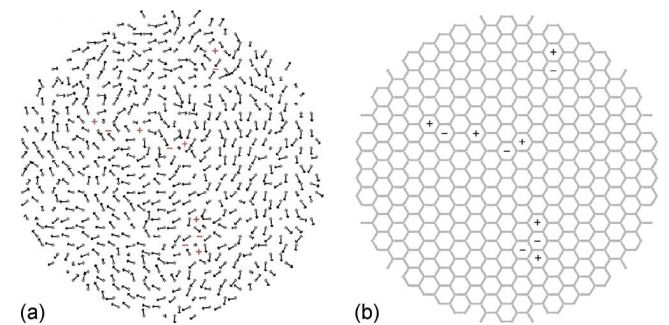

FIG. 6. (Color online) Typical configuration of the vortex-antivortex distribution in a hexagonal lattice at $T=0.7$ for $D=0.1$ and $L=30$. Figure 6(a) shows the spin configuration and the location of the vortices and antivortices. It is possible to see one unpaired vortex and five pairs of vortex-antivortex (plus and minus signs indicate vortex and antivortex, respectively). Figure 6(b) shows only the vortex and antivortex positions inside the corresponding skeleton of the hexagonal lattice of Fig. 6(a).

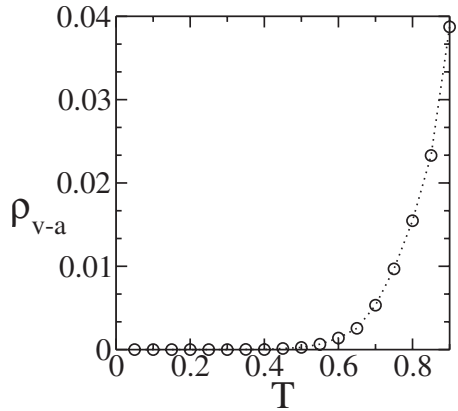

FIG. 7. The pair vortex-antivortex density, $\rho_{v-a}$, as a function of temperature for the same parameters as in Fig. 6.

The same behavior is observed for the square and triangular lattices. At low temperature only one vortex survives in the system. By using spin dynamics we have obtained some preliminary results showing that if the vortex has an out-of-plane component, it can flip at random as temperature increases. It happens in a regime far before the first vortex-antivortex pair appears. The vortex density, $\rho_{v-a}$, is zero up to $T/J=0.45$, however, the vortex can flip as early as at $T/J=0.10$. This observation is in contrast with the reported mechanism of creation-annihilation for switching the vortex core discussed in Refs. 3 and 18. A possible explanation of this result is as follows. The measurements performed in Refs. 3 and 18 were taken at very low temperature, in a regime that even the presence of spin wave excitations are not enough to turn the out-of-plane vortex polarization. When a magnetic pulse is applied, it excites adiabatically the system elevating its temperature beyond the pair creation threshold. At the same time, spin waves are excited. The spin waves excitations can switch the vortex core. We believe that the observed phenomenon can be a fortuitous effect. We also believe that a much more careful simulation has to be done in order to decide about the correct mechanism responsible for the switching of the out-of-plane vortex component. In particular, a rigorous statistical study is of paramount importance.

## IV. CONCLUSION

In this work we investigated, via MC simulation, the conditions for vortex formation in quasi-2D magnetic dots. We used a model Hamiltonian with exchange and dipolar interactions for square, triangular, and hexagonal lattices. Our results showed that a cut-off in the dipolar interaction can give a good approximation only for large dot size and large cut-off radius. Besides that, a finite size scaling, as $D_c = D_0 + 1/A(1+BL^2)$, is proposed to describe the cross over between a capacitor-like state to a vortex state. This behavior is obeyed by the three types of lattices. Inside the vortex phase region it is possible to identify two types of vortices separated by a constant $D=D_c$ line: An in-plane and an out-of-plane vortex. We observed that the out-of-plane phase does not appear for the triangular lattice. In the case of a two layer system we observed that the extra layer of dipoles works as an effective out-of-plane anisotropy, inducing a large z component at the center of the vortex in agreement with the experimental results reported in Ref. 15. We suggest that in a real system, where a multilayer dot is considered, the range where the out-of-plane vortex exists can be considerably large. Also, we analyzed the mechanism responsible for the switching of the out-of-plane vortex component. In contrast to some reported results, we found that the switching mechanism is different from the creation-annihilation vortex antivortex process.

## ACKNOWLEDGMENTS

We are grateful to Dr. L. A. S. Mol for very fruitful discussions. This work was partially supported by CNPq and FAPEMIG (Brazilian Agencies).